\documentclass[12pt,preprint]{aastex}

\newcommand{\Hsi}{\textit{Reuven Ramaty High Energy Solar Spectroscopic Imager}}
\newcommand{\hsi}{\textit{RHESSI}}
\newcommand{\goes}{\textit{GOES}}

\newcommand{\Sdo}{\textit{Solar Dynamics Observatory}}
\newcommand{\sdo}{\textit{SDO}}
\newcommand{\hmi}{Helioseismic and Magnetic Imager}
\newcommand{\hinode}{\textit{Hinode}}

\begin{document}

\title{First Flare-related Rapid Change of Photospheric Magnetic Field Observed by Solar Dynamics Observatory}
\author{Shuo Wang\altaffilmark{1}, Chang Liu\altaffilmark{1}, Rui Liu\altaffilmark{1}, Na Deng\altaffilmark{1}, Yang Liu\altaffilmark{2}, and Haimin Wang\altaffilmark{1}}
\affil{1. Space Weather Research Laboratory, New Jersey Institute of Technology, University Heights, Newark, NJ 07102-1982, USA; haimin.wang@njit.edu}
\affil{2. W.~W. Hansen Experimental Physics Laboratory, Stanford University, Stanford, CA 94305-4085, USA}

\email{*in preparation for formal publication*}

\begin{abstract}
Photospheric magnetic field not only plays important roles in building up free energy and triggering solar eruptions, but also has been observed to change rapidly and permanently responding to the coronal magnetic field restructuring due to coronal transients. The \hmi\ instrument (HMI) on board the newly launched \Sdo\ (\sdo) produces seeing-free full-disk vector magnetograms at consistently high resolution and high cadence, which finally makes possible an unambiguous and comprehensive study of this important back-reaction process. In this study, we present a near disk-center, \goes-class X2.2 flare occurred at NOAA AR 11158 on 2011 February 15 using the magnetic field measurements made by HMI. We obtained the {\it first} solid evidence of an enhancement in the transverse magnetic field at the flaring magnetic polarity inversion line (PIL) by a magnitude of 70\%. This rapid and irreversible field evolution is unequivocally associated with the flare occurrence, with the enhancement area located in between the two chromospheric flare ribbons. Similar findings have been made for another two major flare events observed by \sdo. These results strongly corroborate our previous suggestion that the photospheric magnetic field near the PIL must become more horizontal after eruptions.

In-depth studies will follow to further link the photospheric magnetic field changes with the dynamics of coronal mass ejections, when full Stokes inversion is carried out to generate accurate magnetic field vectors.

\end{abstract}


\section{INTRODUCTION} 
Almost two decades ago, we discovered rapid and permanent changes of vector magnetic fields
associated with flares \citep{wang92,wang94}. Specifically, the transverse field near the flaring magnetic polarity inversion line (PIL) is found to enhance impulsively and substantially, which is also often accompanied by an increase of magnetic shear. Similar trend of photospheric field evolution associated with flares and coronal mass ejections (CMEs) has continued to be observed later on in many observations \citep{wang02b,wang04,wang+liu05,liu05,wang07a,jing08}, and shows some agreement as well when compared to recent model predictions \citep{li10}. Nevertheless, such a study is unavoidably hampered by the obvious limitations of ground-based observation (e.g., seeing variation and the limited number of observing spectrum position), most probably because of which mixed results were also reported \citep{ambastha93,hagyard99,chen94,li00a,li00b}.

On the other hand, flare-related variations in the line-of-sight (LOS) component of photospheric magnetic field have been clearly recognized \citep[e.g.,][]{wang02b,spirock02,yurchyshyn04,sudol05,wang06,wang10,petrie10}. However, it is noted that the changes of LOS field are qualitative in nature without providing an exact understanding of the coronal field restructuring \citep{hudson11}. 

It is notable that systematic vector measurements has been made available with the \hmi\ (HMI) instrument \citep{schou10} on board the newly launched \Sdo\ (\sdo). Its unprecedented observing capabilities give a favorable opportunity to finally resolve any uncertainties regarding the evolution of photospheric magnetic field in relation to flares/CMEs.

In this study, we investigate a near disk-center X2.2 flare on 2011 February 15, where we found the first solid evidence of the enhancement in the transverse field at the flaring PIL using the seeing-free and consistent HMI data. We will discuss the implications of such a change under the context of the recently developed theory.

\section{OBSERVATIONS AND DATA REDUCTION}
The $\beta\gamma\delta$ region NOAA 11158 lies close to the disk center (S21$^{\circ}$, W21$^{\circ}$) when the 2011 February 15 X2.2 flare started at 01:44~UT and peaked at 01:56~UT in \goes\ soft X-ray flux. As a preliminary approach in handling the HMI data, we used the signal of linear polarization as a proxy of the transverse magnetic field strength $B_t$. We normalized the data in a way to keep the $\langle B_t \rangle$ value of a control region (i.e., an $\alpha$-spot in the active region) constant, so that any observed change in the flaring region could be meaningfully reduced. The resulted $B_t$ map has a spatial resolution of 0.5\arcsec\ and a cadence of 12 minutes.

\begin{figure}[t]
\epsscale{.9}
\plotone{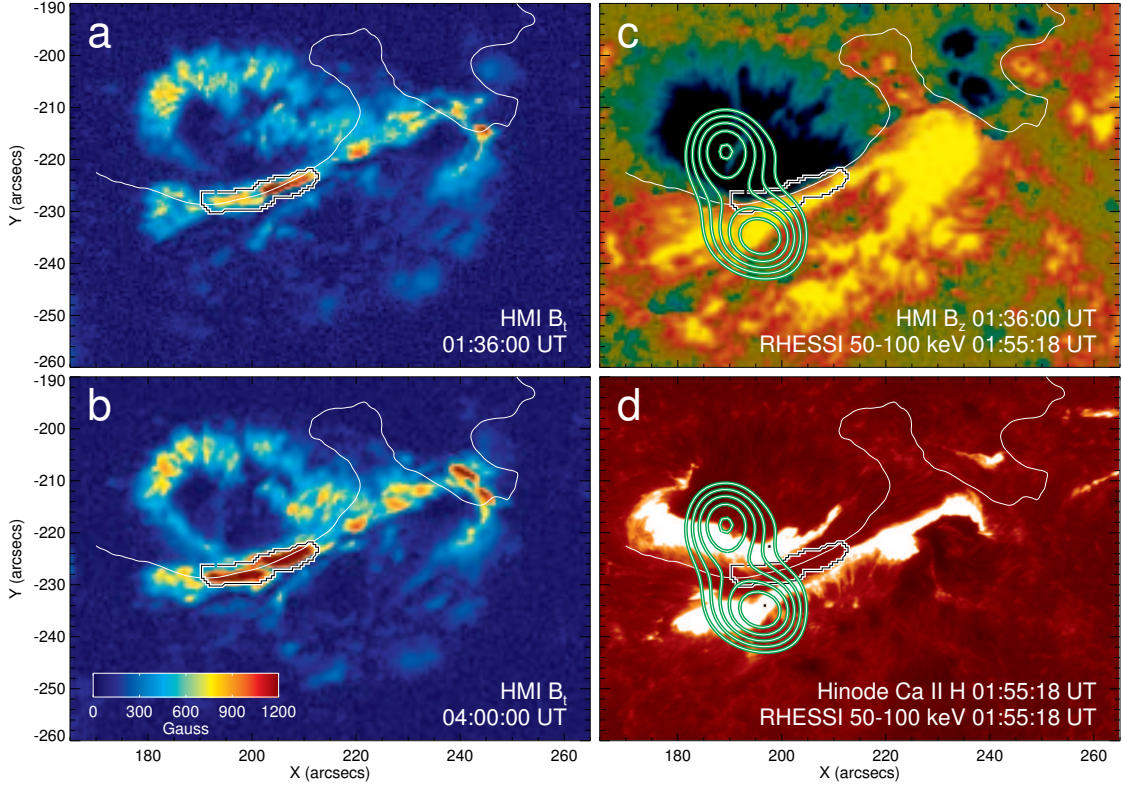}
\caption{Pre- ($a$) and post-flare ($b$) $B_t$ map obtained by HMI clearly showing the enhancement of transverse field at the PIL (white line) in a region enclosed by the white bordered line, which is located directly between the HXR footpoints (green bordered contours) and the portions of the brightest chromospheric ribbons ($c$ and $d$). The CLEAN \hsi\ image is reconstructed for the peak of 50--100~keV emissions from 01:55:00 to 01:55:36~UT obtaining a total count of 1.7~$\times$~10$^4$. The contour levels are 50\%, 60\%, 70\%, 80\%, and 90\% of the maximum flux. \label{f1}}
\end{figure}

The evolution of the flare hard X-ray (HXR) emission was entirely registered by the \Hsi\ \citep[\hsi;][]{lin02}. As a quick assessment of the location of high energy particles bombardment, a CLEAN image \citep{hurford02} was reconstructed using the front segment of detectors 3--8 and natural weighting (giving an FWHM resolution of $\sim$14\arcsec), covering the peak of the 50--100~keV emissions from 01:55:00 to 01:55:36~UT. As a reference, we also present a cotemporal Ca~{\sc ii} H image taken by the Solar Optical Telescope \citep[SOT;][]{tsuneta08} on board \hinode\ \citep{kosugi07}, which shows the chromospheric flare ribbons.

\section{RESULTS}

From the pre- and post-flare $B_t$ maps presented in Figure~\ref{f1}$ab$, a pronounced enhancement in the transverse field strength can be seen in a region at the main flaring PIL (enclosed by the white bordered line). The location of this region is between the two closely spaced chromospheric ribbons, the portions of which are cospatial with the footpoints of the flare HXR emissions in the 50--100~keV (Fig.~\ref{f1}$cd$). 

\begin{figure}[t]
\epsscale{.8}
\plotone{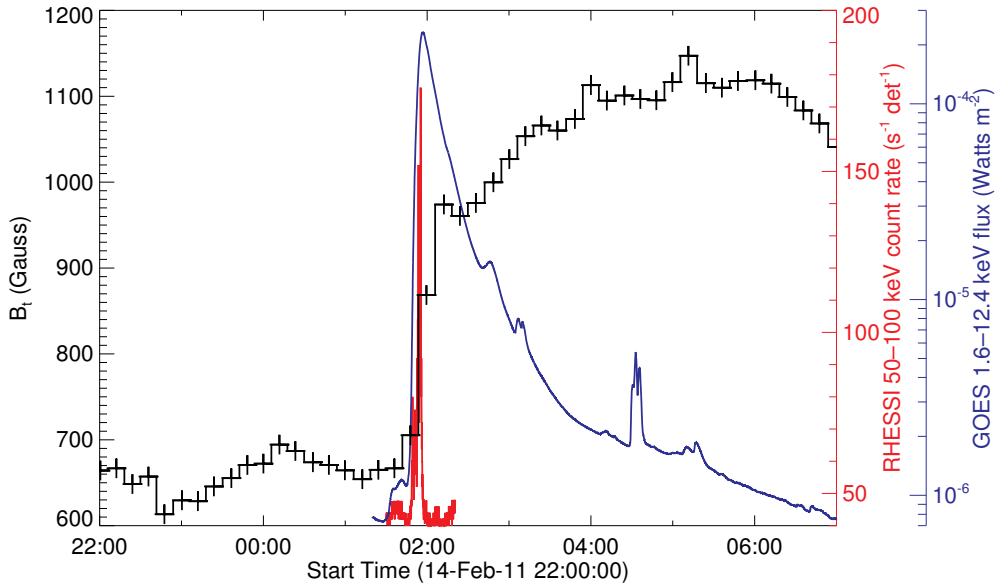}
\caption{Temporal evolution of $\langle B_t \rangle$ of the region enclosed by the white bordered line in Fig.~\ref{f1}, in comparison with the light curves of \hsi\ HXR count rate in the 50--100~keV energy range (red) and \goes\ flux in 1--8~\AA\ (blue). \label{f2}}
\end{figure}

In Figure~\ref{f2}, we plot the temporal evolution of $\langle B_t \rangle$ of this region spanning $\sim$8 hrs centering on the occurrence of the flare. It is evident that the transverse magnetic field increases by $\sim$70\% from $\sim$650~G to $\sim$1100~G, the transition time of which is cotemporal with the peaking of the flare HXR emissions representing the abrupt release of the accumulated free magnetic energy. 

These observations strongly endorse our previous studies, in which similar changes in the transverse field were observed using ground-based data with the influence of seeing \citep[e.g.,][]{wang+liu05,liu05}.

\section{DISCUSSION}
The seeing-free and consistent \sdo/HMI magnetic field observation with unprecedented high spatial and temporal resolution has produced, {\it for the first time}, unambiguous and solid evidence of the rapid, flare-related enhancement in the transverse field at the flaring PIL. Analysis on another two major flares observed by \sdo\ is currently underway, in which we have obtained very similar results. These provide strong support to our previous observations \citep[see][for a review]{wang10}, based on which we have suggested that photospheric magnetic fields respond to coronal field restructuring (namely, an inward collapse after the eruption of CMEs) by tilting toward the surface (i.e., toward a more horizontal state) near the PIL. This view is well in line with the recent theoretical development \citep{hudson08,fisher10}, where the back reaction on the solar surface and interior resulting from the coronal field evolution required to release energy is quantitatively assessed in terms of the downward Lorentz force integrated over the region where the field change is significant.

In the following, we discuss a few points related to observations and modeling work.

\begin{enumerate}

\item The endeavor of studying the flare-related photospheric magnetic field changes has been plagued by the problem that the magnetic field measurement could be affected by flare emissions \citep{qiu03,patterson81}. Here we only concentrate on the magnetic fields at the PIL in between the flare ribbons, where both the flare emission is insignificant and the weak-field approximation is usually acceptable. We anticipate that full Stokes inversion will minimize such effects so that a fully reliable analysis of field evolution can be carried out.

\item Using the magnetic vectors derived from inversion, we will examine the following aspects of magnetic field properties: (1) magnetic shear. It has been demonstrated that increase of magnetic shear and of transverse field strength are interrelated \citep[e.g.,][]{wang02b}. (2) Lorentz force. A correlation between the downward Lorentz force computed from changes in the vector magnetograms and the dynamics of the upward propelled CMEs has been theoretically suggested \citep{fisher10}.

\item The changes of transverse magnetic field can be directly visualized by the evolution of the sunspot penumbral structure. Similar to our previous analysis \citep{wang04a,deng05,liu05,chen07b}, the darkening of penumbrae in the studied region with enhancement of transverse field is quite discernible. However, the decay of penumbrae in the outside boundary of sunspots, which we surmised to be associated with the inward collapse of active region magnetic field, is less obvious in the present event, probably because the source active region is still in the growing phase. As related to the field collapse, we will study both the direct and indirect manifestations, namely, the evolution of longitudinal magnetic field of the entire active region \citep{wang10} and the signatures of coronal field implosion \citep[][and references therein]{liur+wang10}.
\end{enumerate}

\acknowledgments
\sdo\ is a mission for NASA's Living With a Star (LWS) program. \hsi\ is a NASA Small Explorer.

\end{document}